\documentclass[pdflatex,sn-standardnature]{sn-jnl}

\jyear{2026}%

\theoremstyle{thmstyleone}%
\theoremstyle{thmstyletwo}%

\theoremstyle{thmstylethree}%

\DeclareGraphicsExtensions{.eps,.eps.gz,.epsi, .jpg, .png}

\usepackage[superscript,biblabel]{cite}
\usepackage{natbib}
\usepackage{lineno}
\usepackage{xcolor}
\linenumbers

\counterwithin*{figure}{section}
\renewcommand{\refname}{}

\usepackage{newunicodechar}
\newunicodechar{─}{--}
\begin{document}


\title[A massive merger]{\bf \large Evidence of a massive accretion event 1.8 billion years before the Gaia-Sausage-Enceladus merger}

\author*[1]{\small \fnm{Davide} \sur{Massari}}\email{davide.massari@inaf.it}
\author[2]{\small \fnm{Chiara} \sur{Zerbinati}}\email{chiara.zerbinati2@unibo.it }
\author[1]{\small \fnm{Cristiano} \sur{Fanelli}}\email{cristiano.fanelli@inaf.it}
\author[3]{\small \fnm{Amina} \sur{Helmi}}\email{ahelmi@astro.rug.nl}
\author[1,2,3]{\small \fnm{Edoardo} \sur{Ceccarelli}}\email{ceccarelli@astro.rug.nl }
\author[4,5]{\small \fnm{Fernando} \sur{Aguado-Agelet}}\email{faguado@uvigo.gal}
\author[6,7]{\small \fnm{Santi} \sur{Cassisi}}\email{santi.cassisi@inaf.it}
\author[3,8]{\small \fnm{Ewoud} \sur{Wempe}}\email{ewoud.wempe@obspm.fr}
\author[6,9]{\small \fnm{Matteo} \sur{Monelli}}\email{matteo.monelli@inaf.it}
\author[10]{\small \fnm{Andrea} \sur{Bellini}}\email{bellini@stsci.edu}
\author[3]{\small \fnm{Thomas} \sur{Callingham}}\email{t.m.callingham@astro.rug.nl }
\author[3]{\small \fnm{Hanneke C.} \sur{Woudenberg}}\email{woudenberg@astro.rug.nl }
\author[11]{\small \fnm{Roger} \sur{Cohen$^{11}$}}\email{rc1273@physics.rutgers.edu}
\author[12]{\small \fnm{Carme} \sur{Gallart}}\email{carme.gallart@iac.es}
\author[13]{\small \fnm{Elena} \sur{Pancino}}\email{elena.pancino@inaf.it}
\author[13]{\small \fnm{Sara} \sur{Saracino}}\email{sara.saracino@inaf.it}
\author[1,14]{\small \fnm{Maurizio} \sur{Salaris}}\email{M.Salaris@ljmu.ac.uk}
\author[2]{\small \fnm{Alessio} \sur{Mucciarelli}}\email{alessio.mucciarelli2@unibo.it}


\affil*[\footnotesize 1]{\footnotesize \orgdiv{Osservatorio di Astrofisica e Scienza dello Spazio di Bologna}, \orgname{INAF}, \orgaddress{\street{Via Gobetti 93/3}, \city{Bologna}, \postcode{I-40129}, \country{Italy}}}

\affil[\footnotesize 2]{\footnotesize \orgdiv{Dipartimento di Fisica e Astronomia}, \orgname{Universit\`a degli Studi di Bologna}, \orgaddress{\street{Via Piero Gobetti 93/2}, \city{Bologna}, \postcode{I-40129}, \country{Italy}}}

\affil[\footnotesize 3]{\footnotesize \orgdiv{Kapteyn Astronomical Institute}, \orgname{University of Groningen}, \orgaddress{\street{Landleven 12}, \city{Groningen}, \postcode{9747 AD}, 
\country{The Netherlands }}}

\affil[\footnotesize 4]{\footnotesize \orgdiv{Escola de Enxe\~{n}ar\'ia de Telecomunicaci\'on}, \orgname{Universidade de Vigo}, \orgaddress{\city{Vigo}, \postcode{36310}, 
\country{Spain}}}

\affil[\footnotesize 5]{\footnotesize 
\orgname{Universidad de La Laguna}, \orgaddress{\street{Avda. Astrof\'isico Fco. S\'anchez}, \city{La Laguna}, \postcode{E-38205}, \state{Tenerife}, \country{Spain}}}

\affil[\footnotesize 6]{\footnotesize \orgdiv{Osservatorio Astronomico di Abruzzo}, \orgname{INAF}, \orgaddress{\street{Via M. Maggini}, \city{Teramo}, \postcode{64100}, 
\country{Italy}}}

\affil[\footnotesize 7]{\footnotesize \orgdiv{INFN - Sezione di Pisa}, \orgname{Universit\'a di Pisa}, \orgaddress{\street{Largo Pontecorvo 3}, \city{Pisa}, \postcode{56127}, 
\country{Italy}}}

\affil[\footnotesize 8]{\footnotesize \orgdiv{LIRA}, \orgname{Observatoire de Paris, Université PSL, Sorbonne Université, Université Paris Cité, CY Cergy Paris Université}, \orgaddress{\street{CNRS}, \city{Meudon}, \postcode{92190}, 
\country{France}}}

\affil[\footnotesize 9]{\footnotesize \orgdiv{Osservatorio Astronomico di Roma}, \orgname{INAF}, \orgaddress{\street{Via Frascati 33}, \city{Monte Porzio Catone}, \postcode{00078}, 
\country{Italy}}}

\affil[\footnotesize 10]{\footnotesize \orgname{Space Telescope Science Institute}, \orgaddress{\street{3700 San Martin Drive}, \city{Baltimore}, \postcode{21218}, \state{MD}, \country{USA}}}

\affil[\footnotesize 11]{\footnotesize \orgdiv{Department of Physics and Astronomy}, \orgname{Rutgers the State University of New Jersey}, \orgaddress{\street{136 Frelinghuysen Road}, \city{Piscataway}, \postcode{08854}, \state{NJ}, \country{USA}}}

\affil[\footnotesize 12]{\footnotesize \orgname{Instituto de Astrof\'isica de Canarias}, \orgaddress{\street{Calle V\'ia L\'actea s/n}, \city{La Laguna}, \postcode{E-38206}, 
\country{Spain}}}

\affil[\footnotesize 13]{\footnotesize \orgdiv{Osservatorio Astrofisico di Arcetri}, \orgname{INAF}, \orgaddress{\street{Largo E. Fermi 5}, \city{Firenze}, \postcode{50125}, 
\country{Italy}}}

\affil[\footnotesize 14]{\footnotesize \orgdiv{Astrophysics Research Institute}, \orgname{Liverpool John Moores University}, \orgaddress{\street{146 Brownlow Hill}, \city{Liverpool}, \postcode{L3 5RF}, 
\country{UK}}}

\maketitle

\newpage
{\bf The merger history of the Galaxy has been traced back firmly to redshift 2 (10 Billion years ago). While evidence for the existence of at least one more significant merger before this time has been presented, its interpretation is yet to be fully established. Here we show that the population of globular clusters around the Galaxy depicts three distinct age-metallicity sequences, one associated with the progenitor of the Milky Way, one with the merger with Gaia-Enceladus 10 billion years ago, and a third intermediate sequence associated to at least one merger which we estimate took place about 1.8 billion years before Gaia-Enceladus. This discovery has been possible thanks to exquisite {\it Hubble Space Telescope} data and sophisticated analysis that enables very precise relative age determination of globular clusters. The newly identified sequence reveals that this merger took place with an object of stellar mass similar to that of Gaia-Enceladus ($\simeq5\times10^8$ M$_{\odot}$), and which deposited most of its mass in the inner 6 kpc of the Milky Way. The identification of a third merger event in the inner Galaxy puts to rest earlier debates, and honoring previous works we name the progenitor system Low-energy-Kraken-Heracles, or LKH for short.}

According to the currently favored
cosmological framework, large galaxies like the Milky Way (MW) have grown through time, at least in part, via mergers with smaller systems. Our knowledge of the history of events that contributed to the assembly of our Galaxy has greatly improved\cite{helmi20} thanks to the advent of the {\it Gaia} mission \cite{prusti16} and large spectroscopic surveys (e.g., APOGEE \cite{majewski17}, GALAH \cite{buder21}). The most recent merger, still ongoing, is the one with the Sagittarius dwarf galaxy \cite{ibata94}, which fell in about 6 billion years (Gyr) ago \cite{ruizlara20}. Prior to this, the most significant merger experienced by the MW known was with the Gaia-Sausage-Enceladus (GSE) dwarf galaxy \cite{helmi18, belokurov18}, and ended about $10$ Gyr ago \cite{gallart19, montalban21}. Due to the high mass ratio of this event (about 1:4-1:5) this led to the development of the thick disk as we see it today, both through dynamical heating and significant star formation \cite{gallart19, dimatteo19,  belokurov20}. A few other accretion events have also taken place in between these two significant mergers (i.e. between 6 and 10 Gyr ago), all involving smaller dwarf galaxies \cite{myeong19, naidu20, ruizlara22}. 

Reconstructing what happened in the earliest phases of the MW's life, more than 10 Gyr ago, is significantly more challenging. As we proceed back in time, the mass of our Galaxy must have been smaller and more comparable to that of the accreting building blocks \cite{horta24}. Because of this and the extremely short evolutionary timescales, the properties of the infant MW and those of the merging systems could be difficult to distinguish either dynamically or chemically \cite{santistevan20, orkney22, garciabethencourt23}. The very short mixing timescales (driven in part by chaos in the inner Galaxy induced by the rotating bar) imply that some of the dynamical memory must have been  erased\cite{woudenberg25}. Furthermore, because of the intricacies of modeling chemical evolution, state-of-the-art cosmological hydrodynamical simulations currently provide limited guidance on all the chemical elements that could potentially be used to distinguish the early building blocks from one another\cite{santistevan20, garciabethencourt23}.

The current understanding is that the earliest phases of the MW evolution are represented by dynamically hot \cite{belokurov22, chandra24}, chemically-unevolved \cite{malhan24, horta25} stellar populations, concentrated in the inner 3 kpc of the present-day Galaxy \cite{rix22}. These populations are likely a mix of stars born in the main progenitor ({\it in-situ}), and of stars born in building-blocks accreted early on, but the dispute about their origin is not settled yet. Attempts to observationally isolate the {\it in-situ} population from the accreted building blocks have yielded debated results so far, with cross-contamination being the major challenge. Using a compilation of literature age estimates, a group of globular clusters (GCs) at low orbital energy (the ``Low-energy" group, hereafter), concentrated within 6 kpc from the Galactic centre, was identified and presumed to be related to an accretion event \cite{massari19}. Subsequently, these GCs were interpreted to be associated with the Kraken merger event \cite{kruijssen20}, an entity that had been predicted\cite{kruijssen19} using cosmological numerical simulations on the basis of the number of GCs with unknown origin along the branch of the age-metallicity relation populated by accreted systems. However, the accreted origin of these GCs has been contested on the basis of chemo-dynamical arguments \cite{belokurov24, chen24}. Similarly, the search for accreted stars in the inner Galaxy led to the identification of a metal-poor and chemically-unevolved population, that was termed Heracles \cite{horta21}. However, also its origin is debated\cite{belokurov22, vasini24}, since the stars' chemical properties are not exclusive of accreted populations and may be found in early {\it in-situ} populations as well (at least in the elements considered thus far and in the metal-poor regime). A more recent analysis\cite{horta25} including the study of the spatial distribution of the Heracles stars, nonetheless seems to provide further support to the presence of an accreted population. 

The difficulty to untangle the {\it in-situ} population from that accreted in the earliest phases of the MW evolution\cite{horta25, rix24}, suggests that more information is needed. Crucially, chronological information can be used to sequence the events that took place in the proto-Galaxy. 
While relative ages are difficult to determine accurately and precisely for individual stars, this is possible for star clusters\cite{massari23, vandenberg13}.

Here we homogeneously analyze for the first time, a sample of 17 globular clusters (GCs) in the inner Galaxy, where the debris of the earliest accretion events should be located \cite{pfeffer20}. These GCs have been observed with the {\it Hubble} Space Telescope (HST) in the F606W and F814W optical bands, and 15 of these have been dynamically associated with the ``Low-energy" group in previous works \cite{massari19, callingham22}. Using our recently developed highly sophisticated analysis, we are able to determine extremely precise relative GC ages with typical errors of a few hundred Myr\cite{massari23} (see the Methods Section 1 for details). The increase in precision is due to the minimization of 
systematic errors, driven in previous work by the heterogeneity of datasets and methodologies, as well as by a more sophisticated treatment of each cluster's photometry.

With our analysis of these 17 GCs, this brings the total number of GCs with homogeneous age determination to  39\cite{massari23, aguadoagelet25, ceccarelli25, massari25}. Besides the Low-energy group, we expect\cite{massari25note} this sample to also contain in-situ GCs and some associated to GSE.  
To robustly determine the most likely number of progenitors required to describe this chrono-dynamical data set, we fit a Bayesian model with multiple components (each representing a progenitor system) using 
the dynamic nested sampling package {\tt dynesty}.  We assume each progenitor to follow an age-metallicity relation (AMR), described using a simple chemical evolution model\cite{prantzos08} (see the Methods Section 2 for details). We also assume the associated GCs to be clustered in their Jacobi energy ($E_J$), (average) vertical action $J_z$ and (average) circularity $\eta= L_z/L_z(E)$ (see the Methods section for a justification of this choice of dynamical quantities). To compute these, we have used a Galactic mass model including a rotating bar\cite{woudenberg25, mcmillan17}, and computed the averages over 200 orbits. The distribution of clusters is taken to be a mixture of $K$ Student-t distributions, where $K$ is to be determined, and would correspond to the number of progenitors.   
This statistical approach allows us to identify distinct chrono-dynamical groups while properly accounting for observational uncertainties and model evidence.
By adopting flat priors on the AMR characteristic parameters, we estimated the Bayesian evidence ($\log Z$) for the different cases having $K$ = 2, 3 and 4 separate progenitors. We find that the case with the highest evidence is the one that admits 3 progenitors. 
In comparison to it, the 2-progenitor case has $\Delta\log Z=5.3$, and is therefore decisively disfavored according to both the scales by Jeffreys and by Kass \& Raftery. The 4-progenitors case instead has $\Delta\log Z=2.2$, and is thus strongly and moderately disfavoured, respectively, according to the two scales.

\begin{figure}[htp!]
\centering
\includegraphics[width=\textwidth]{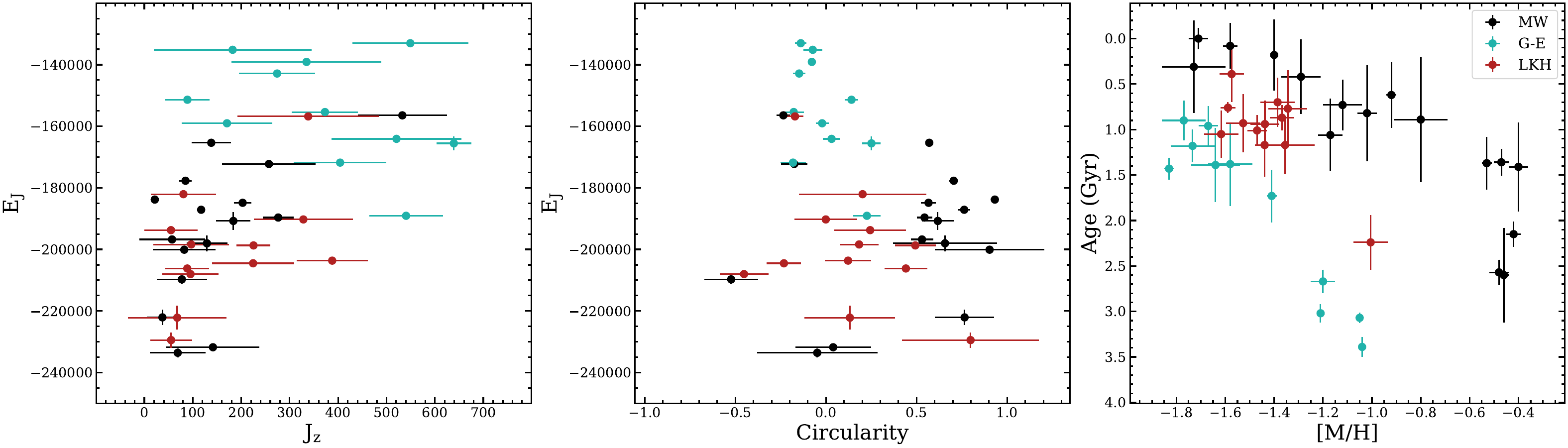}
\caption{\small {\bf Chrono-dynamical properties of our sample of 39 GCs.} \emph{Left panel}: distribution in the E$_J$ vs J$_Z$ dynamical space of the 39 GCs with homogeneous age estimate (filled symbols, see the Methods Section). GCs are color-coded according to their association (with membership probability $>50$\%) with GSE (cyan symbols), the MW (black symbols) or the LKH (red symbols), as determined by our Bayesian chrono/dynamical analysis. Uncertainties are derived as the 16th- and the 84th-  percentile of each parameter distribution computed over 200 orbits. \emph{Middle panel}: same in the E$_J$ vs circularity space. \emph{Right panel}: distribution in the age-metallicity plane of the three populations of GCs. The relative age scale along the y-axis is defined by setting the age resulting from the isochrone fit of the oldest GC (t$_{iso}=14.48$ Gyr) as zero.}
\label{fig:fig1}
\end{figure}

Fig.~\ref{fig:fig1} shows the distribution of the 39 GCs in our sample in the dynamical spaces (left and middle panels) and in the age-metallicity plane (right panel), colour-coded by their membership as determined by our statistical analysis in the most favoured case with 3 components. It is immediately evident that there is an intermediate population of GCs (red symbols) that follows an age-metallicity sequence that is well separated from that traced by GSE GCs (cyan symbols) and the {\it in-situ} GCs (black symbols). These 12 GCs (out of the 15 in our sample initially associated to the ``low-energy" group) are confined within the inner 6 kpc from the Galactic centre. Hence, most of the ``low-energy" GCs must have formed in an environment that is neither GSE, nor the Milky Way, and given their  distinct dynamical properties, it follows that this population of GCs is necessarily part of an independent merger event.

\begin{figure*}[htp!]
\centering
\includegraphics[width=0.355\textwidth]{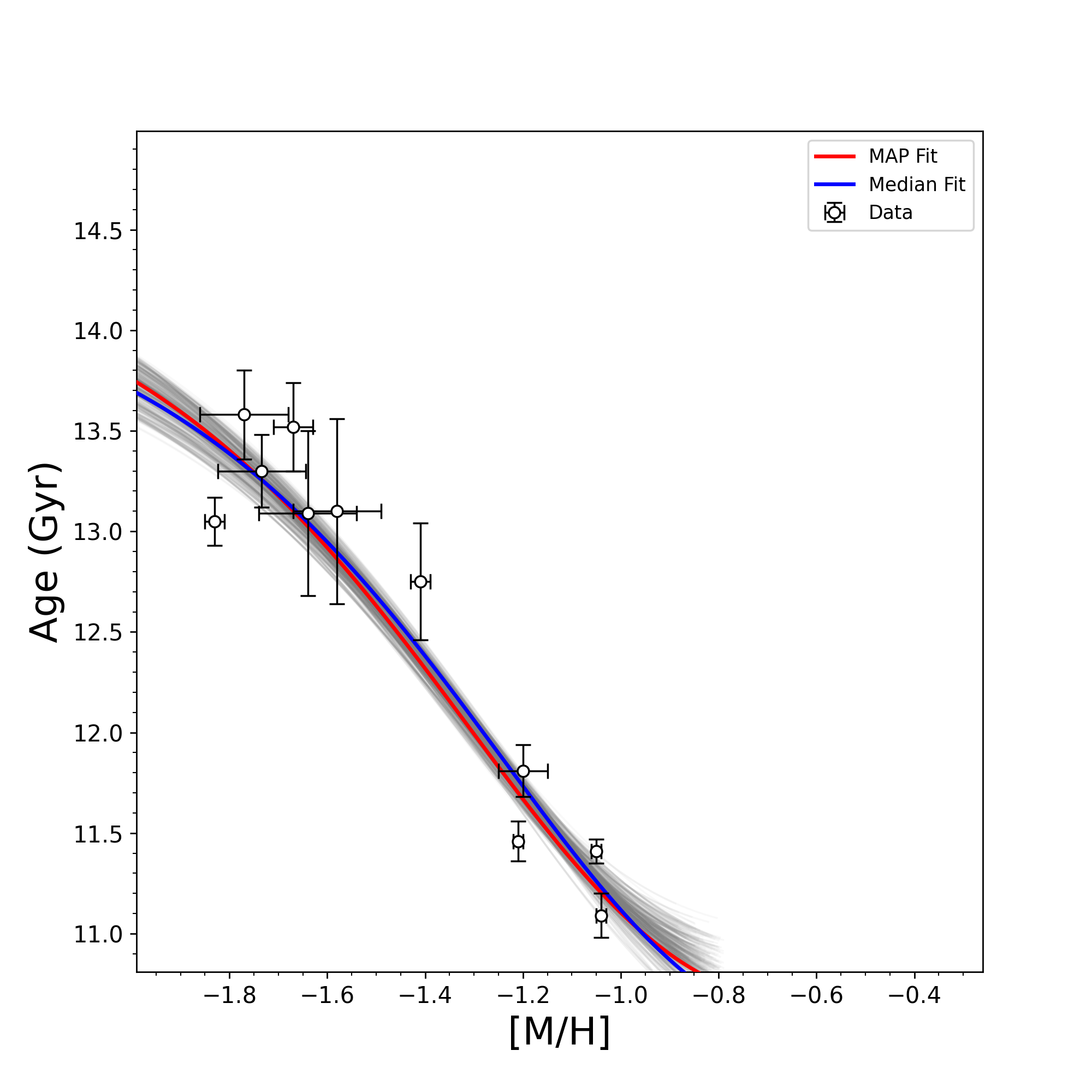}
\hspace{-6mm}
\includegraphics[width=0.355\textwidth]{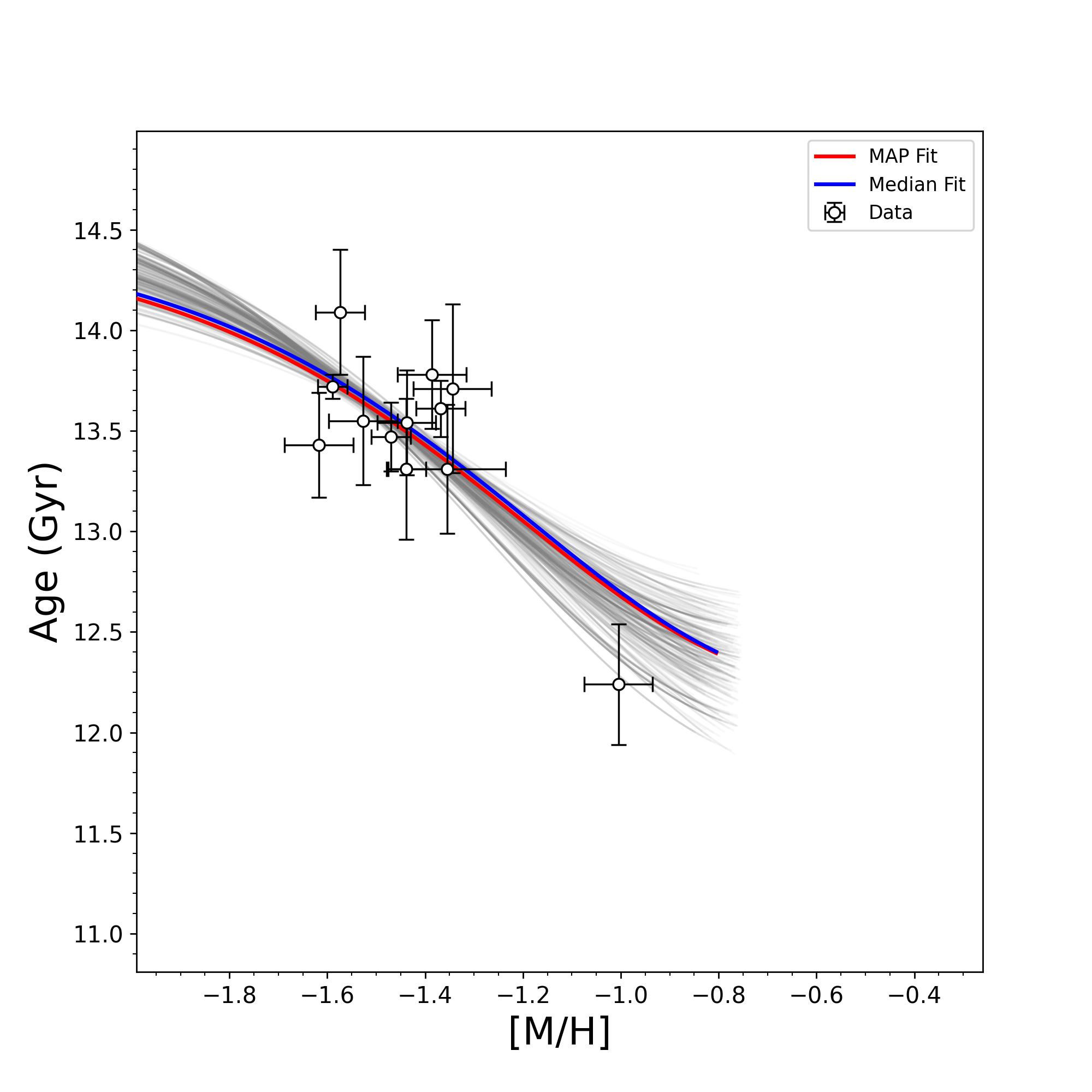}
\hspace{-6mm}
\includegraphics[width=0.355\textwidth]{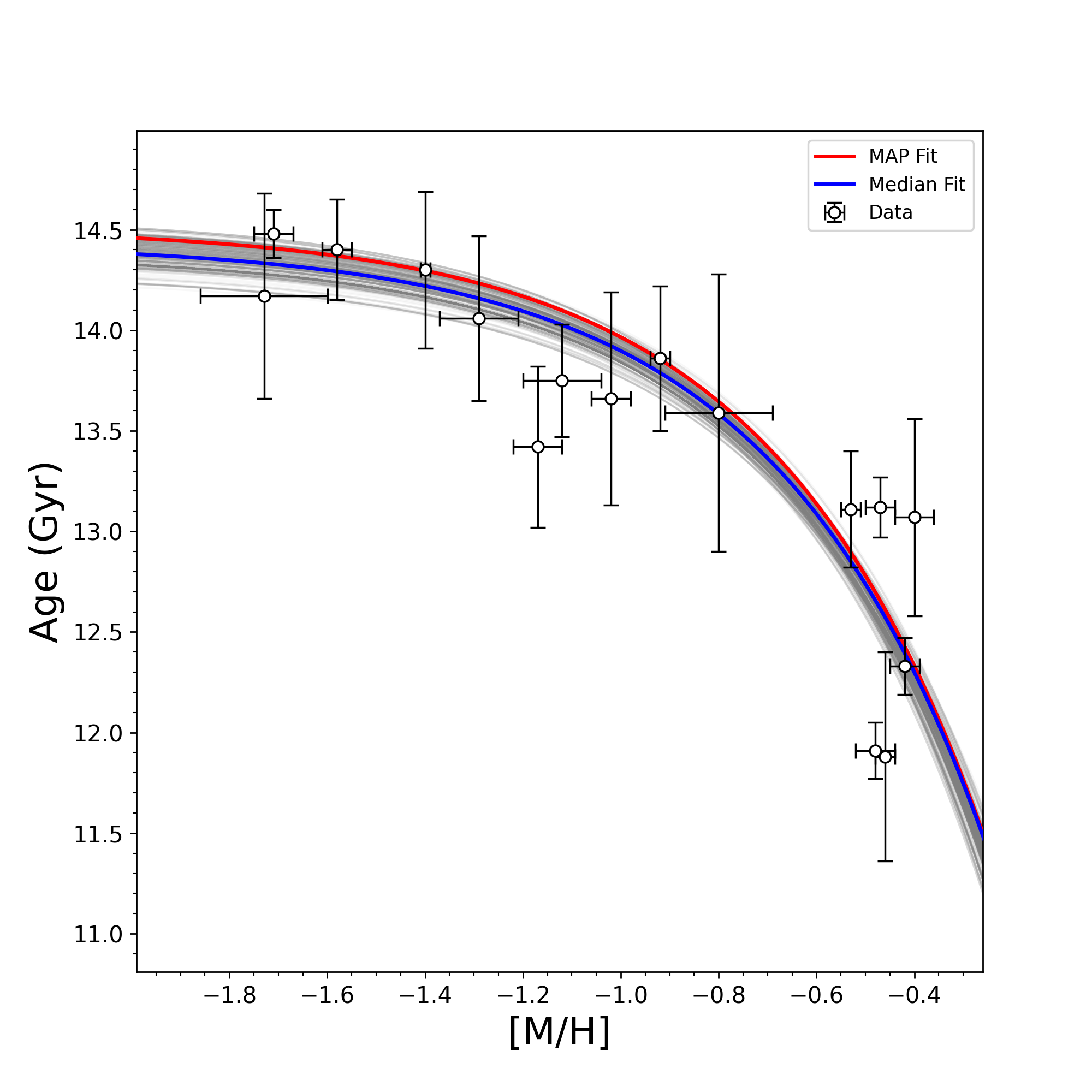}
\caption{\small {\bf AMR models fit.} \emph{Left-hand panel}:  fit to the GSE GCs. \emph{Middle panel}:  fit to LKH GCs. \emph{Right-hand panel}:  fit to the {\it in-situ} MW GCs. The individual GC members of the three samples are selected as described in the Methods Section. The red line corresponds to the fit with the Maximum A Posteriori (MAP) value, while the blue line represents the median fit.}
\label{fig:fig2}
\end{figure*}

While the above statistical analysis allows us to unveil the presence of three  distinct populations of GCs (see the Methods section 3 for details), to more accurately constrain the properties of their progenitor systems, we fit the AMR of the three groups separately, using the GCs with a high-probability ($p > 50$\%) of membership. As before we use a simple chemical evolution model \cite{prantzos08}, with parameters depending on the stellar mass of a galaxy \cite{dekel03}, the time at which star formation started, and the time at which it ended, that we take to represent the galaxy's accretion time (see Methods for details). Fig.~\ref{fig:fig2} shows the individual fits of the three AMRs. From the comparison between GSE and the low-energy GCs, we find that the merger event identified here has a stellar mass consistent with that of GSE within the uncertainties. Moreover, its accretion time is $\sim1.8$ Gyr earlier than GSE. These relative measurements are robust, and are anchored to absolute values by taking as a reference for GSE\cite{helmi18} a stellar mass of $\sim5\times10^8$ M$_{\odot}$ and an accretion time of t$_{accr}\simeq10.5$ Gyr \cite{fernandezalvar25}. 
It follows that we have unveiled a very significant merger event that took place $\simeq12.3$ Gyr ago (at redshift $z>4$), with a mass ratio that must have been higher than that of the GSE event\cite{helmi18, gallart19}, i.e.$> 0.2 - 0.3$. 
The estimated stellar mass for these two progenitors is in agreement with the host stellar mass - GC metallicity relation\cite{brodie06}, according to which a galaxy this massive (with absolute magnitude in B-band\cite{savaglio05} M$_{B}\simeq-17.5$) should host a GC system of mean metallicity [M/H]$\simeq-1.4$.

This early and massive merger has characteristics that could link it to the putative accreted Heracles stellar population, which is also restricted to the inner Galaxy\cite{horta21}. While this region of the Galaxy likely contains debris from all the mergers that took place in the early stages of formation of the Milky Way, our identification may be seen as the cleanest selection extending to low metallicity of the most massive contributor to the Heracles' accreted sub-population. Nonetheless, we cannot rule out the presence of other (presumably less massive) progenitors without GCs or whose GCs have been fully disrupted.  Since the members of this progenitor stem mostly from the Low-energy group\cite{massari19}, later associated with the putative Kraken\cite{kruijssen20}, we name the progenitor Low-energy-Kraken-Heracles, or LKH. 

This is the first time that a significant event occurred during the early evolution of the proto-Galaxy has been identified from the intricate sequence of mergers and {\it in-situ} star formation, and thereby settles the debates mentioned earlier. Moreover, our study establishes the AMR as currently the only tracer able to distinguish in the metal-poor regime ([Fe/H]$\lesssim-1.6$) the main progenitor of our Galaxy, namely the building block that evolved at the highest rate. 
 
\begin{figure}[htp!]
\centering
\includegraphics[scale=0.35]{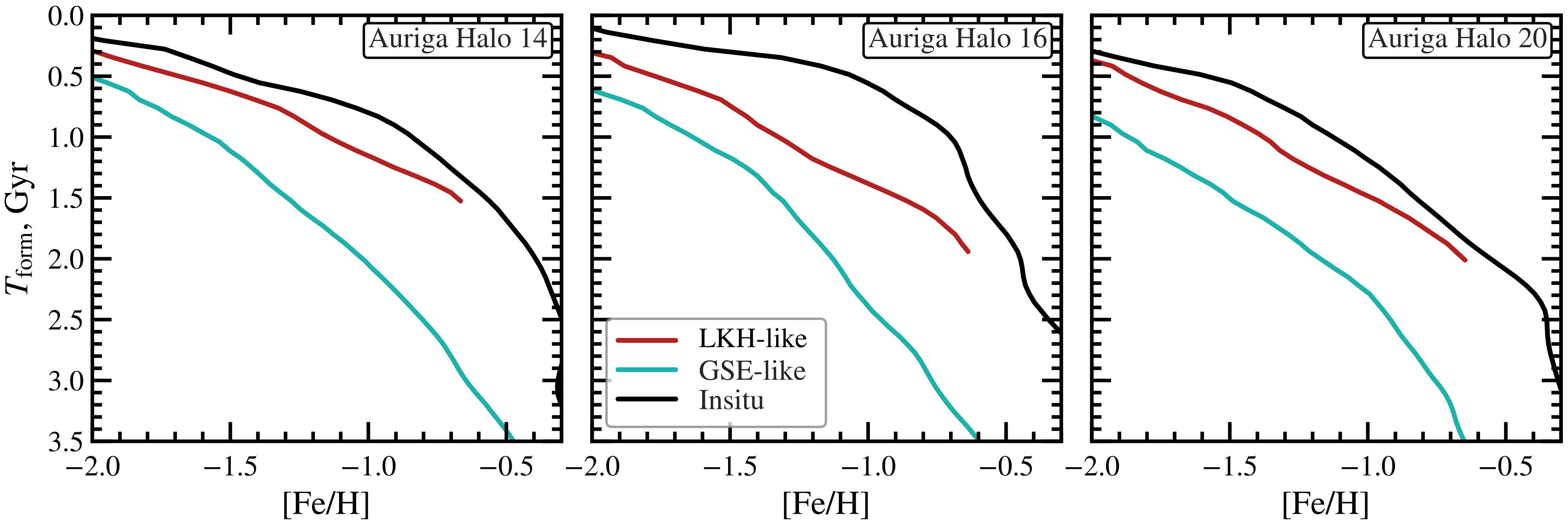}
\caption{\small {\bf AMR in simulated MW analogs.} Median trend (solid lines) for the AMR of the main progenitor (black), a GSE-like merger (cyan) and a LKH-like event (red) in three MW analogs of the Auriga Level-4 simulations.}
\label{fig:amr_simul}
\end{figure}

Our conclusions are strongly supported by cosmological simulations, which typically predict between 1 and 4 significant mergers in the history of a Milky Way-like galaxy \cite{Robertson2005}. Furthermore, the three AMRs identified in our work are in fact remarkably well reproduced by several Milky Way analogs of the Auriga simulations\cite{Auriga}, as shown in Fig.~\ref{fig:amr_simul}. 
In the three simulated galaxies represented in the figure (Au14, Au16 and Au20), the median AMR defined by the stars are shown for the main progenitor, a GSE-like merger (defined as the largest accretion event that occurred between 2.5 Gyr and 5 Gyr from the start of the simulation) and a LKH-like event (defined as the largest event in the first 2.5 Gyr).
The AMRs in the simulations follow a global trend that is in good agreement with what is observed, in particular showing a clear separation among the AMR of the three progenitors by a few hundreds Myr at any metallicity, that matches qualitatively the separation observed with the GCs analyzed in this paper. 
Future precise stellar ages for large samples of stars in the inner MW, particularly measurements coming from asteroseismology missions such as PLATO\cite{rauer25}, or derived with CMD-fitting techniques using {\it Gaia} data will be key to unravel the series of events that took place in the first Gyr of the MW life with even finer detail.

\newpage
\bibliography{test}

\clearpage
\noindent\textbf{\large Methods}
\\

\noindent{\bf 1. Age estimates~}\\

The relative age of all the clusters analyzed in this work have been determined with the strictly homogeneous method developed within the Cluster Ages to Reconstruct the Milky Way Assembly (CARMA) project \cite{massari23, aguadoagelet25, ceccarelli25, niederhofer25}. All the details have been extensively described in paper I of the series. Here we remark that the precision we achieve is due to the minimization of the sources of systematic errors, which in turn is achieved through the adoption, for {\it all} the CARMA GCs, of $i$) homogeneous HST photometry in the F606W and F814W bands, corrected for differential reddening \cite{milone12} and decontaminated from field stars via proper-motion-based membership \cite{nardiello18}; $ii$) the most up-to-date stellar evolution models from the Bag of Stellar Tracks and Isochrones (BaSTI) database \cite{pietrinferni24, hidalgo18} and $iii$) the same Markov chain Monte Carlo (MCMC)-based isochrone fitting techinque, which provides at the same time an estimate for age, global metallicity [M/H], distance and color-excess E(B-V) on the same homogeneous scale. We remark that our age estimates should not be interpreted on an absolute scale, because it is well known that every set of theoretical models has a different absolute scale based on the adopted input physics. In this sense, GC ages older than the age of the Universe should not raise concern. They should be interpreted bearing in mind that they could be inaccurate, but highly precise, which is what matters for all the results presented here.

The number of GCs for which we determine an age estimate in the CARMA framework for the first time here is 17, namely: 15 originally associated with the ``low-energy" group and 2 dynamically interpreted as {\it in-situ}. The photometry for most of these GCs comes from the public database of the HUGS project \cite{piotto15, nardiello18}. The astro-photometric catalogs of NGC~6256, NGC~6401 and NGC~6453 are taken from another source\cite{cohen21}. Finally, the photometry of NGC~4372 has been obtained with the same codes and methods used for the other GCs \cite{bellini10}, by working on the images from GO-15982 (PI: Reinhart).

\begin{figure*}[!th]
   \centering
   \begin{minipage}{0.45\textwidth}
        \centering
        \includegraphics[width=1.0\textwidth]{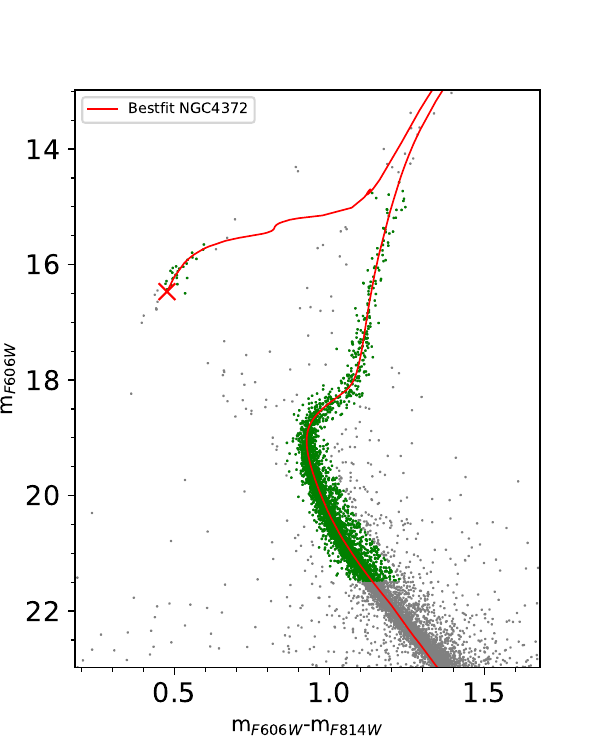}
        \includegraphics[width=1.0\textwidth]{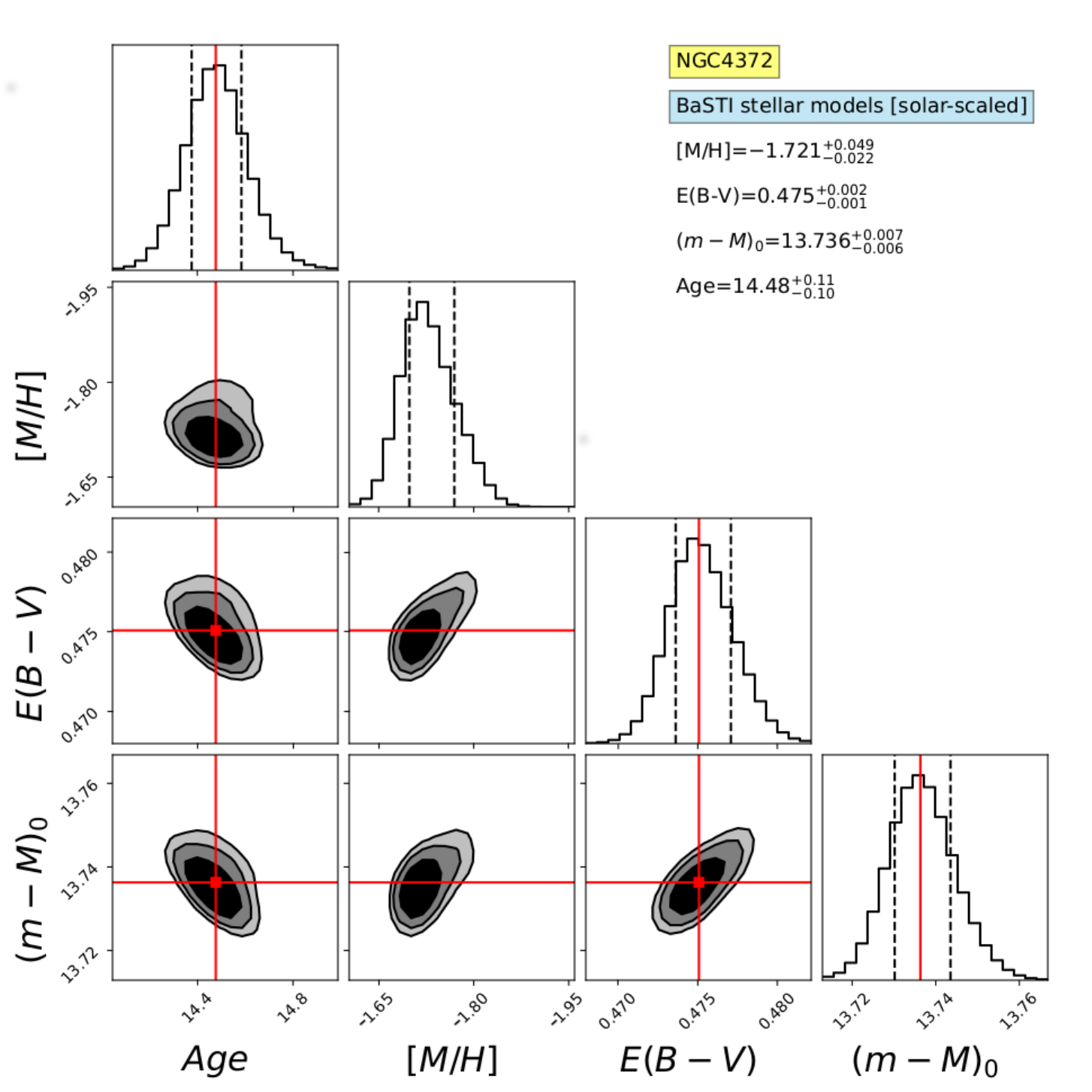}
   \end{minipage}\hfill
   \begin{minipage}{0.45\textwidth}
        \centering
        \includegraphics[width=1.0\textwidth]{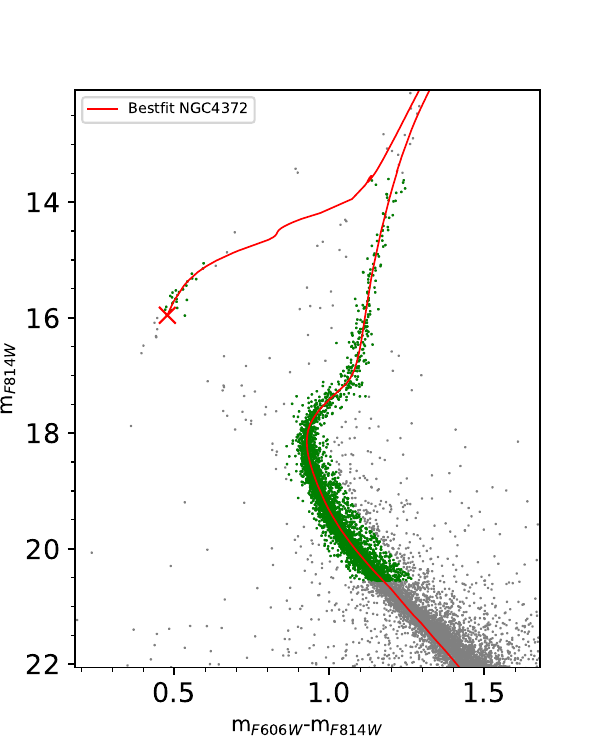}
        \includegraphics[width=1.0\textwidth]{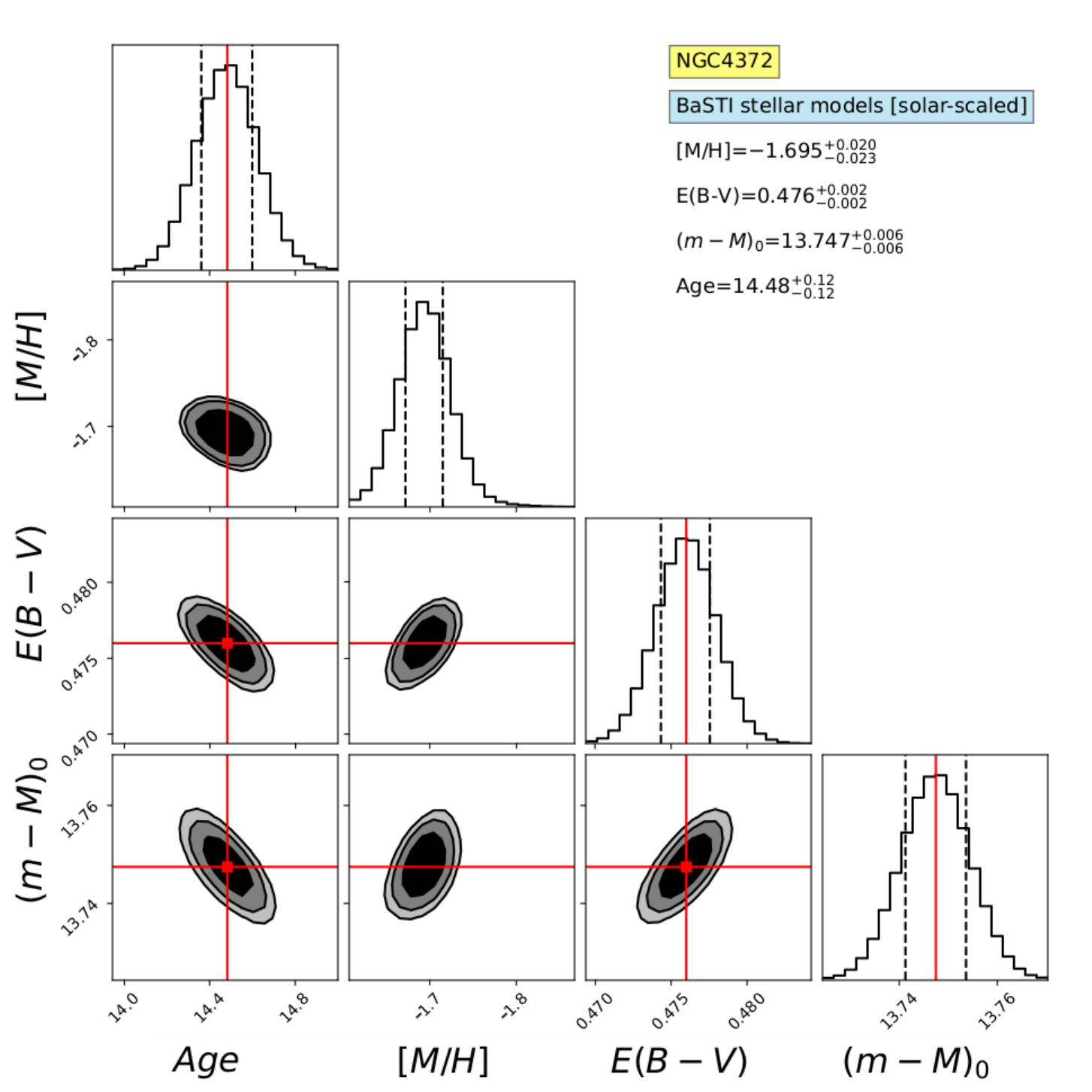}
   \end{minipage}
   \caption{{\bf Results of the isochrone fit for NGC~4372.} Top left panel: best fit model in the ($\it{m_{\mathrm{F606W}}}$, $\it{m_{\mathrm{F606W}}} - \it{m_{\mathrm{F814W}}}$) CMD. Top right panel: best fit model in the ($\it{m_{\mathrm{F814W}}}$, $\it{m_{\mathrm{F606W}}} - \it{m_{\mathrm{F814W}}}$) CMD. Bottom left panel: posterior distributions for the output parameters and the best-fit solution, quoted in the labels, in the ($\it{m_{\mathrm{F606W}}}$, $\it{m_{\mathrm{F606W}}} - \it{m_{\mathrm{F814W}}}$) CMD. Bottom right panel: posterior distributions for the output parameters and the best-fit solution, quoted in the labels, in the ($\it{m_{\mathrm{F814W}}}$, $\it{m_{\mathrm{F606W}}} - \it{m_{\mathrm{F814W}}}$) CMD.}
              \label{fig:ages_4372}%
    \end{figure*}

\begin{figure*}[h]
 \centering
 \includegraphics[width=0.45\linewidth]{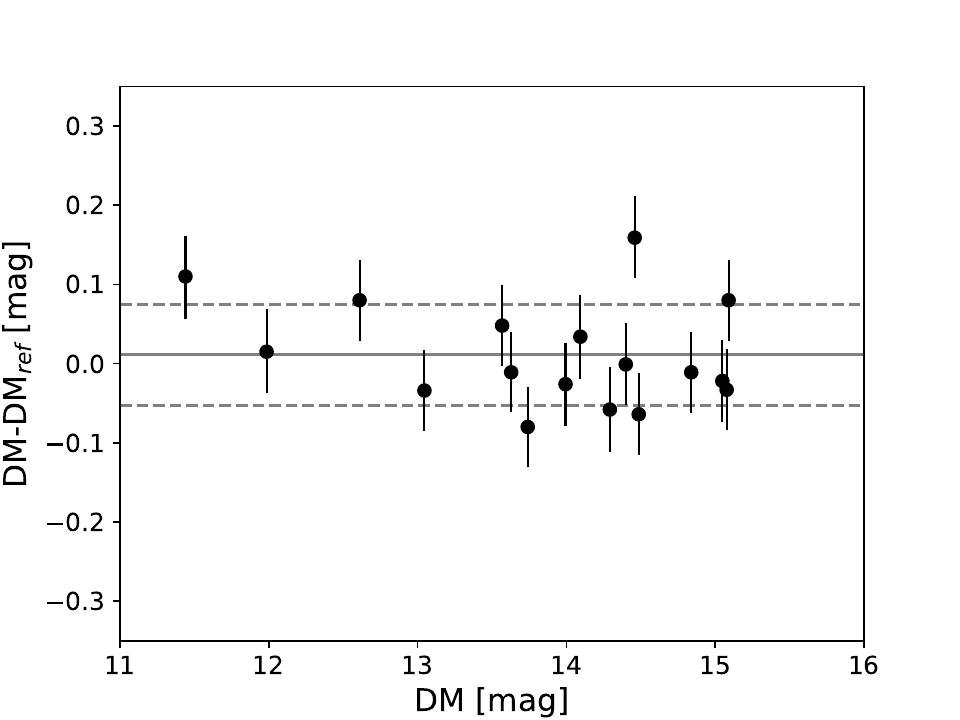}
 \includegraphics[width=0.45\linewidth]{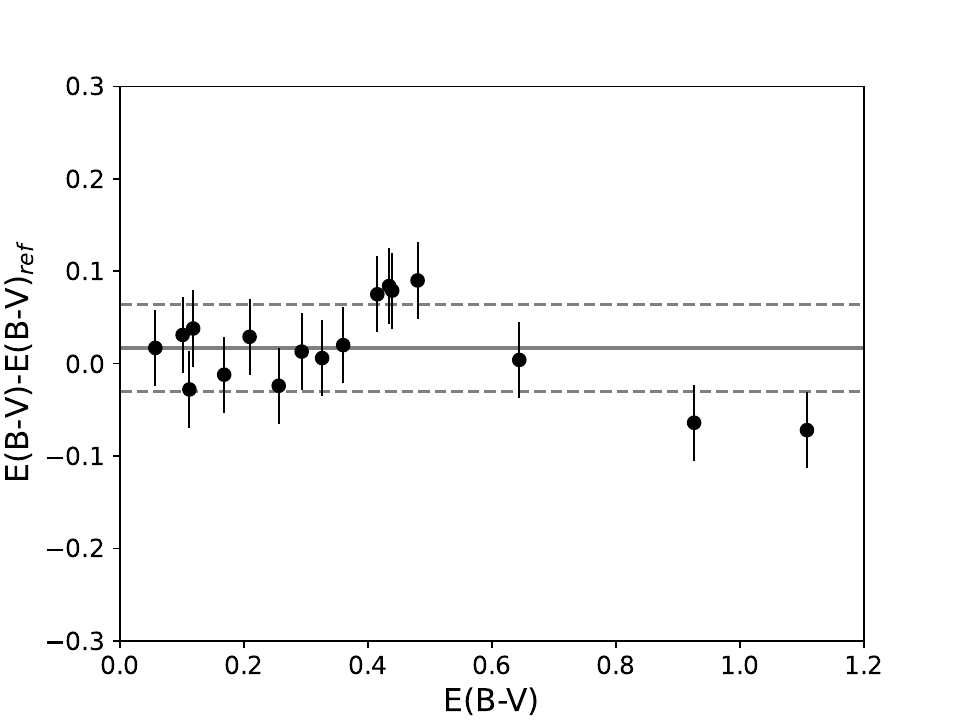}
 \includegraphics[width=0.45\linewidth]{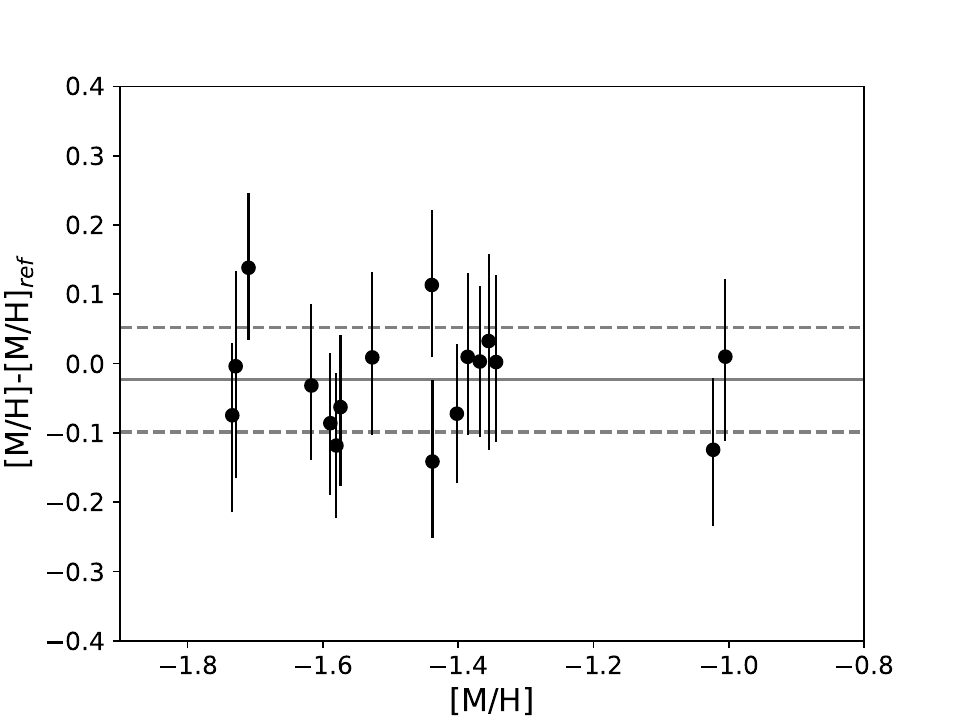}
 \caption{ {\bf Comparison with the literature.} Differences between the output of our isochrone fit and literature values for distance modulus, color-excess and global metallicity. The mean difference is shown with a solid grey line, whereas the 1$\sigma$ dispersion around the mean is indicated with dashed grey lines.}
 \label{fig:liter_check}
\end{figure*}

The result of the isochrone fit is summarized in Table~\ref{tab:iso}. An example of an individual fit is presented in Fig~\ref{fig:ages_4372}. As a sanity check, we compare the values of the free parameters that we obtained from the isochrone fit with existing literature estimates. The comparison is shown in Fig.~\ref{fig:liter_check}, and demonstrates that the mean difference (solid grey lines) is always consistent with zero within 1$\sigma$ (dashed grey lines). Literature color-excesses E(B-V)$_{ref}$ and distance moduli DM$_{ref}$ have been taken from the Harris catalog \cite{harris96}, but the same good consistency is found when considering distance estimates by other works\cite{baumgardt21}. The global metallicity [M/H]$_{ref}$ has been computed starting from the [Fe/H] values of the Harris catalog, transformed according to the equation\cite{salaris93}: [M/H] = [Fe/H]+$\log(0.694\times10^{[\alpha/{\rm Fe}]}+0.306)$, and using a relation \cite{aguadoagelet25} between [Fe/H] and [$\alpha$/Fe].

Significant effort has been devoted to the determination of the relative age of Milky Way GCs in the past\cite{chaboyer96, marinfranch09}.
To compare the uncertainty on our age estimates with previous works, we use the samples by Dotter et al.\cite{dotter10} and by VandenBerg et al.\cite{vandenberg13}, as they determine GC relative ages using the photometry in the same F606W and F814W HST filters adopted here. Since the precision on the GC age depends on GC properties such as distance or reddening, we only compare GCs in common with the two works. Of the 17 GCs for which we provide a CARMA age for the first time here, 12 are in common with the Dotter sample, and 12 (not strictly the same ones) with the VandenBerg sample. Our uncertainties are on average a factor of 3.5 better than the former (a mean error of 0.26 Gyr vs 0.91 Gyr) and a factor 1.6 better than the latter (a mean of 0.26 Gyr vs a mean of 0.43 Gyr). While some of the sources of errors affecting ours and these works estimates should in principle be similar, the most notable differences are the following:
($i$) our CMDs are always corrected for differential reddening and decontaminated from field stars using proper motions measurements; ($ii$) whenever a GCs is known to host stellar populations enriched in iron, helium (Y) or C+N+O (which affect also the photometry in optical bands, contrarily to the standard multiple population phenomenon), we remove the chemically peculiar stars using HST UV photometry; ($iii$) we correct our theoretical models for temperature-dependent extinction; ($iv$) we do not fix the GCs metallicity, distance and reddening, but we leave them as free parameters of the fit.\\

\noindent{\bf 2. AMR models}\\

The simple chemical evolution model\cite{prantzos08} enables inferring an age-metallicity relation under relatively straightforward assumptions. For a constant star-formation rate (SFR), the age $t$ is related to the metallicity as
\begin{equation}\label{eq:model}
    t(x) = t_f + \Delta t \cdot \exp\left[-\frac{10^x}{p}\right]
\end{equation}
where  $x$ = [M/H] is the metallicity, $p$ is the effective chemical yield (in units of $Z_\odot$), and $\Delta t=t_i - t_f$ denotes that star formation started $t_i$ Gyr ago and ended after a time interval $t_i - t_f$, with  $t_i > t_f$.  In the case of an exponentially declining SFR, i.e. $\propto \exp(-\beta t)$, the age and metallicity are related through 
\begin{equation}\label{eq:sfr}
    t(x) = t_{i} - 10^{x}/(\beta~p)
\end{equation}
when the star formation timescale is much longer (i.e. $t_i \gg t_f)$. In the opposite case, it can be shown that Eq.~\ref{eq:sfr} reduces to Eq.~\ref{eq:model}. For the MW, which is the progenitor that endured star formation the longest, we will adopt an exponentially declining SFR, hence Eq.~\ref{eq:sfr}. For the accreted systems instead, since both $\beta$ and $t$ are small, we adopt Eq.~\ref{eq:model}. We remark that in our analysis we adopt $t_f$ as the indicator for the merger accretion time. Although such an assumption is not novel\cite{kruijssen20}, it has been shown that a delay of 0.5-1.5 Gyr might occur between the end of GC formation and the conclusion of an accretion event \cite{hughes19}. However, this does not affect our reconstructed merger timeline, as it relies on the relative comparison between the $t_f$ values inferred for each different progenitor.\\

We assume the effective chemical yield $p$ to be constant over time and to be proportional to the stellar mass of the progenitor \cite{dekel03} according to the relation \cite{prantzos08}
\begin{equation}\label{eq:p}
    p = 0.005 \left(\frac{M}{10^6\ \mathrm{M}_{\odot}}\right)^{0.4}.
\end{equation}
The value of the coefficient in Eq.\ref{eq:p}, was determined to better represent the yield of the progenitor systems of the MW, where SNIa did not have the time to contribute significantly\cite{prantzos08}. As such, it is a factor 2 smaller than the original value describing present-day dwarf galaxies\cite{dekel03}.\\

\noindent{\bf  3. Statistical analysis and characterization of the progenitors}\\ 

As described in the main paper, we determine the most likely number of progenitors in our dataset using chrono-dynamical information\cite{callingham22} and employing the dynamic nested sampling package {\tt dynesty}. This is done in a step-wise procedure as follows:\\
$i$) to anchor the AMR of the {\it in-situ} and GSE progenitors, we first selected a few clusters with homogeneous chemo-chrono-dynamical properties that in the literature are consistently and unambiguously associated with either of these progenitors. We assign NGC~5927, NGC~6304, NGC~6352, NGC~6388, NGC~6441, NGC~6496 (according to previous work\cite{massari23, carretta22}), and NGC~6218, NGC~288, ESO452-11 and NGC~6362 (based on previous analysis\cite{ceccarelli25}) to the {\it in-situ} component, whereas NGC~362 and NGC~1261 were assigned to the GSE component\cite{ceccarelli25}.\\
$ii$) We assume that each of the multiple components/progenitors  follows an AMR and is clustered in the dynamical space described by the separable parameters $(E_J$, $J_Z$, $\eta$). $E_J$ is preferred to the more often used $E$, because the latter is not an integral of motion for many of the clusters in our sample because of the Galactic bar. Furthermore, we prefer to replace the traditional $L_z$ by the circularity $\eta$ because $E_J$ is already related to $L_z$ ($E_J = E - \Omega_b L_z$).  Furthermore, $\eta$ contains information both on the eccentricity of the orbit as well as on its sense of rotation. Finally $J_z$ is sensitive to complementary information (as is $L_\perp$) on an  orbit, and although, like $\eta$ it is not fully conserved, it is an action and we do account for its variation in our analysis.

Concerning the AMR, we use flat priors on the progenitor mass $M_\star$, $t_i$, $t_f$ and (whenever necessary) $\beta$, enforcing the constraint $t_i-t_f>1$ Gyr. We do not sample directly the yield parameter $p$ but obtain it from $M_\star$ using Eq.~\ref{eq:p}, which therefore implies a non-flat prior on $p$. Each of the dynamical parameters are modeled as a one-dimensional Student-t likelihood ($\nu=4$) centered on the mean value of each component, and with a scale given by the sum in quadrature of the measurement uncertainties and the intrinsic dispersion. Clustering is implemented directly in the total likelihood through a $K$-component mixture.\\
$iii$) We compute the Bayesian evidence for the cases with 2, 3 and 4 components. 

We find that the most likely case is that including three progenitors. We tested that this result is robust to the choice of the GCs anchored in step $i)$. In fact, by reducing these to the six most metal-rich ones, the Bayesian evidence continues to favor $K=3$ over $K=2$ decisively ($\Delta\log Z=12.9$), and over $K=4$ more moderately ($\Delta\log Z=2.1$). At the same time, our Bayesian mixture analysis provides the values for the AMR parameters, as well as the posterior membership probabilities for each component. This allows us to define the samples of GCs belonging to each progenitors as those having a probability membership $>50\%$: 
\begin{itemize}
    \item {\it in-situ} component: other than those assigned initially, the analysis added NGC~4372, NGC~6205, NGC~6397, NGC~6401, NGC~6656, and NGC~6752 ;
    \item GSE component: already including NGC~362 and NGC~1261, further encompasses NGC~1851, NGC~2298, NGC~2808, NGC~6341, NGC~5897, NGC~6779, NGC~6809, NGC~7089, and NGC~7099;
    \item LKH, the third component is associated with:  NGC~4833, NGC~5286, NGC~5986, NGC~6093, NGC~6121, NGC~6144, NGC~6254, NGC~6256, NGC~6453, NGC~6535, NGC~6541, and NGC~6681. 
\end{itemize}
    
We note that there is significant overlap between the original low-energy group and LKH, as 91\% of its GCs (10 out of 11) are in common\cite{massari25note}. Moreover, we tested that these associations are robust to the inclusion of a small (200 Myr) intrinsic spread to the individual AMRs. Additional or improved age measurements might cause some GCs associations to change, especially those of the GCs that are at the boundaries of the three populations in the chrono/dynamical space. We have, however, tested via bootstrapping, that the existence of the three AMR sequences is robust to such changes, as is the derivation of their properties.

We use this list of likely members of each progenitor to characterize these better. Although our Bayesian mixture model has fitted AMRs, it uses all clusters simultaneously and hence provides less strict constraints on their properties. Therefore to characterize each progenitor in terms of stellar mass $M_{\star}$ (through $p$) and accretion time $t_i$, we MCMC fit each individual AMR using priors informed by the 5D nested-sampling posterior. For $M_{\star}$ we adopt lognormal priors, while for $t_i$, $t_f$, and (whenever applicable) $\beta$, we use wide flat priors.
Specifically, for the mass of GSE we adopt $\ln M_{\star} \sim \mathcal{N}(\mu,\sigma^2)$ with $(\mu,\sigma)=(20,1)$, corresponding to a median $M_{\star}\approx 5\times10^8\,M_{\odot}$ and a $1\sigma$ scatter of almost a factor of three around the median. For LKH we assume the same prior motivated by the 5D nested-sampling posterior, which favors a mass of a similar order of magnitude.
The star-formation times $t_i$ and $t_f$ are regulated by uniform priors within fixed bounds. In particular, we use $t_i \in [t_{i,\min}, t_{f,\max}]$ and $t_f \in [t_{f,\min}, t_{f,\max}]$, with $(t_{i,\min},t_{f,\min},t_{f,\max})=(8,8,15)\,$Gyr. When sampled (linear AMR), $\beta$ is assigned a wide uniform prior within fixed bounds $\in $[0.01,2.00]. 
The likelihood function is assumed to be Gaussian in both metallicity and age, and is defined as:
\begin{equation}\scriptsize
\ln \mathcal{L}=
\sum_{n=1}^{N} \ln \mathcal{N}\!\left(z^{\rm obs}_n \mid z^{\rm true}_n,\sigma_{z,n}^2\right)
+
\sum_{n=1}^{N} \ln \mathcal{N}\!\left(t^{\rm obs}_n \mid t^{\rm model}_n,\sigma_{t,n}^2+\sigma_{\rm int}^2\right)
\end{equation}
where $z^{\rm obs}_n$ and $t^{\rm obs}_n$ are the observed metallicities and ages, with measurement uncertainties $\sigma_{z,n}$ and $\sigma_{t,n}$, respectively, and $z^{\rm true}_n$ are latent "true" metallicities sampled in the MCMC. The model prediction $t^{\rm model}_n$ is given by the adopted AMR parametrization evaluated at $z^{\rm true}_n$ and parameters $(p,t_i,t_f)$ (or $(p,t_i,\beta)$ for the exponentially decaying SFR form). Here $\ln\mathcal{N}(x\mid\mu,\sigma^2)$ denotes the full Gaussian log-density. The age variance includes an additional intrinsic-scatter term, added in quadrature to the observational uncertainty, for which we adopt a weakly informative prior. 

For each dataset we run an MCMC sampler with $n_{\rm walkers}=128$, performing $n_{\rm steps}=20000$ iterations and discarding the first $25\%$ as burn‐in. We initialize all the walkers with scattered starting points to ensure broad coverage of the parameter space. The mass scaling parameter of each walker is set by uniformly drawing a test mass along its previous range and mapping it to $p$, while the formation times $t_i$ and $t_f$ are offset from their lower bounds by small Gaussian perturbations. We assessed the MCMC convergence using the Gelman–Rubin ($\rm \hat{R}$) diagnostic, effective sample size (ESS), and autocorrelation analysis. All the parameters yielded $\rm \hat{R}$ values $<$1.01, an ESS $\gt 10^6$, and a first‐lag autocorrelations $\le \lvert0.04\rvert $, proving good chain mixing and highly efficient sampling.

\begin{figure}
    \centering
    \includegraphics[width=1\linewidth]{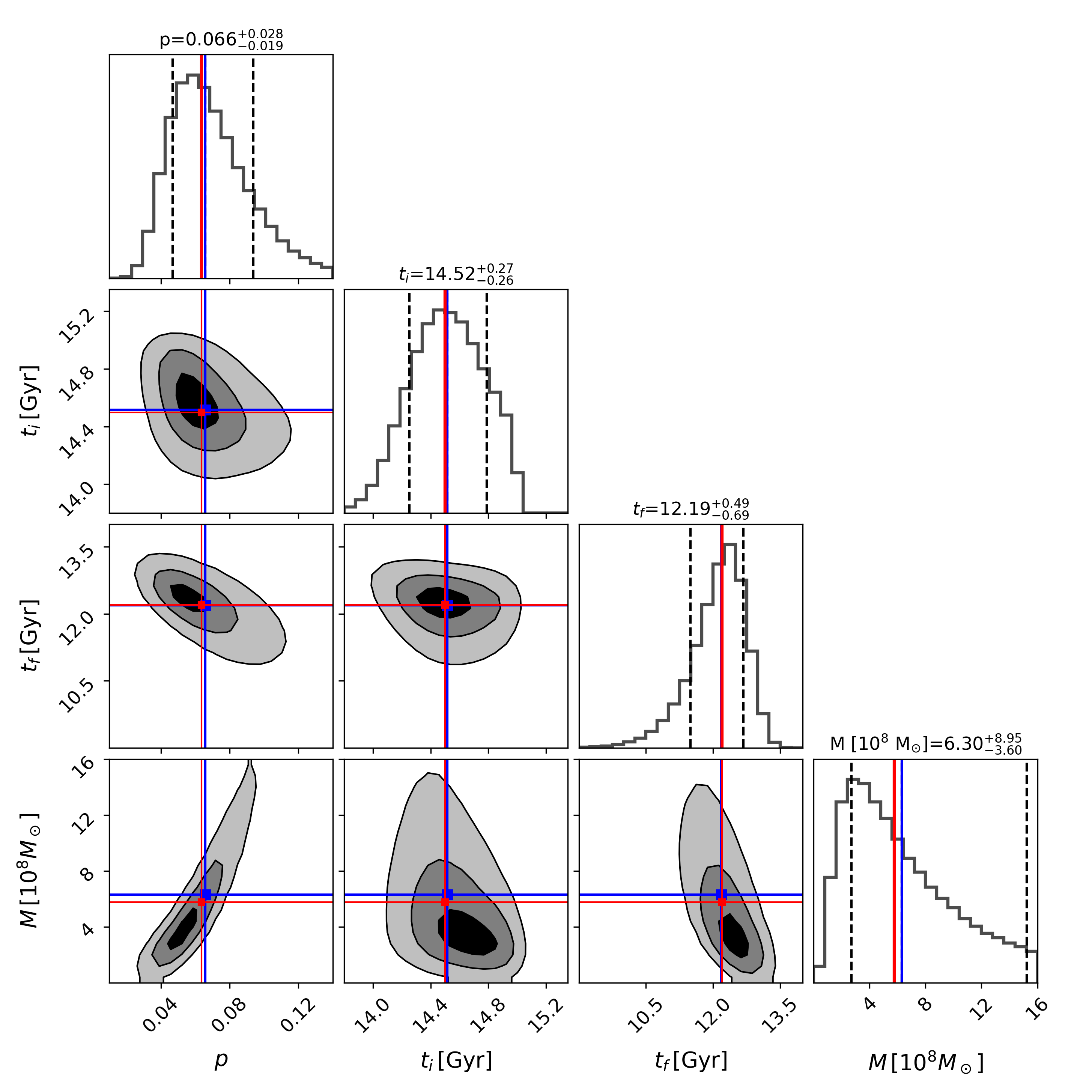}
    \caption{{\bf Corner plot of the LKH AMR model MCMC fit.} Corner plot of the joint posterior for the four global parameters, yield p, star formation initial time $t_i$, accretion time $t_f$, and derived mass M (in units of 10$^8$ M$_\odot$), for the LKH GCs. 
    Along the diagonal are the 1D histograms with dashed lines marking the 16th and 84th percentiles. Solid blue lines indicate the median and solid red lines the MAP, with titles giving the median $\pm$ uncertainties. Off-diagonal panels show the 2D posterior contours at the 16th, 50th and 84th percentile levels, with overplotted blue and red lines denoting the medians and MAP estimates for each parameter pair.
}
    \label{cornerplot}
\end{figure}

\begin{figure}
    \centering
    \includegraphics[width=1\linewidth]{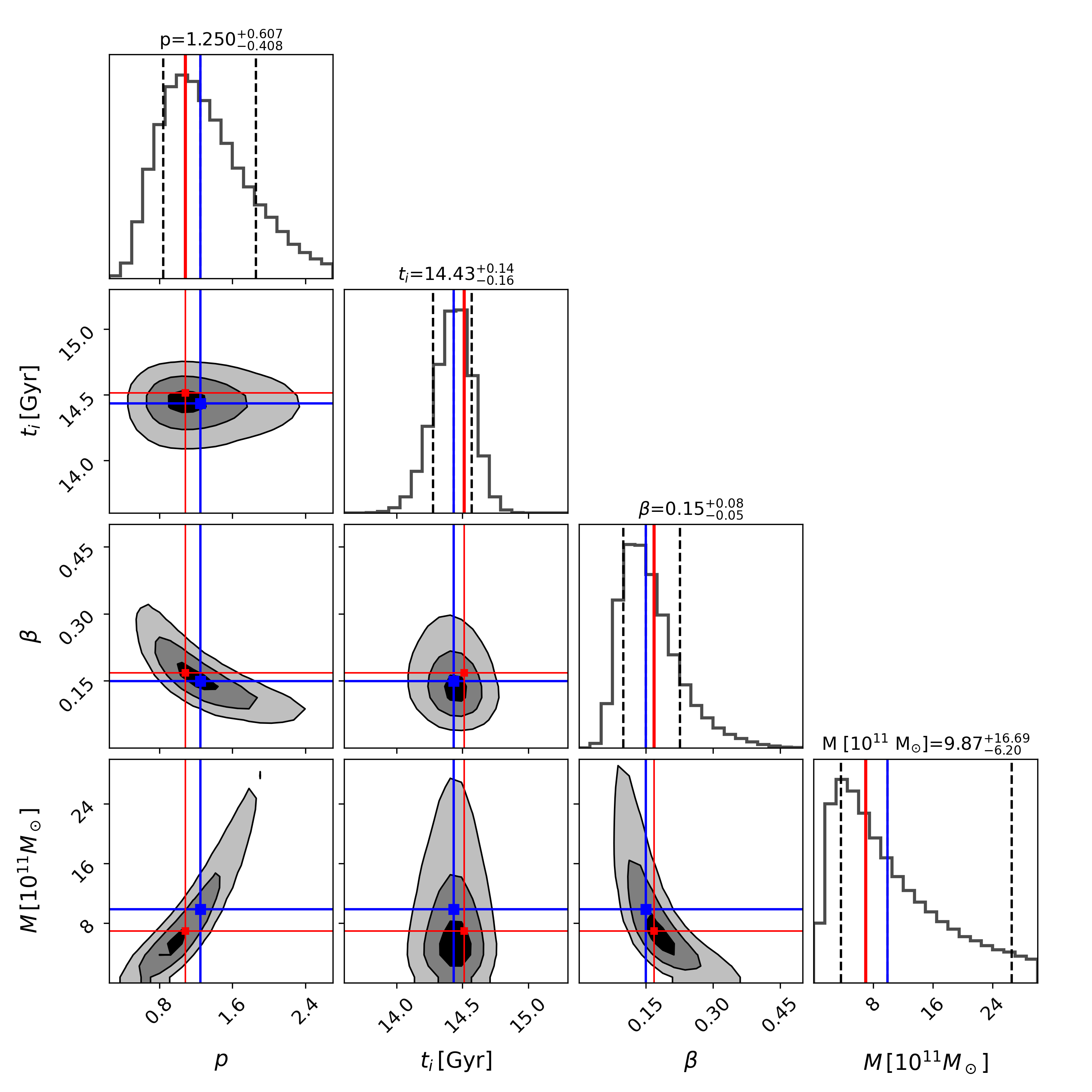}
    \caption{{\bf Corner plot of the MW AMR model MCMC fit.} Corner plot of the joint posterior for the four global parameters, yield p, star formation initial time $t_i$, coefficient of the exponential declination of the SFR $\beta$, and derived mass M (in units of 10$^{11}$ M$_\odot$), for the MW GCs. 
    Along the diagonal are the 1D histograms with dashed lines marking the 16th and 84th percentiles. Solid blue lines indicate the median and solid red lines the MAP, with titles giving the median ± uncertainties. Off-diagonal panels show the 2D posterior contours at the 16th, 50th and 84th percentile levels, with overplotted blue and red lines denoting the medians and MAP estimates for each parameter pair.
}
    \label{MWcornerplot}
\end{figure}

\begin{figure}
    \centering
    \includegraphics[width=1\linewidth]{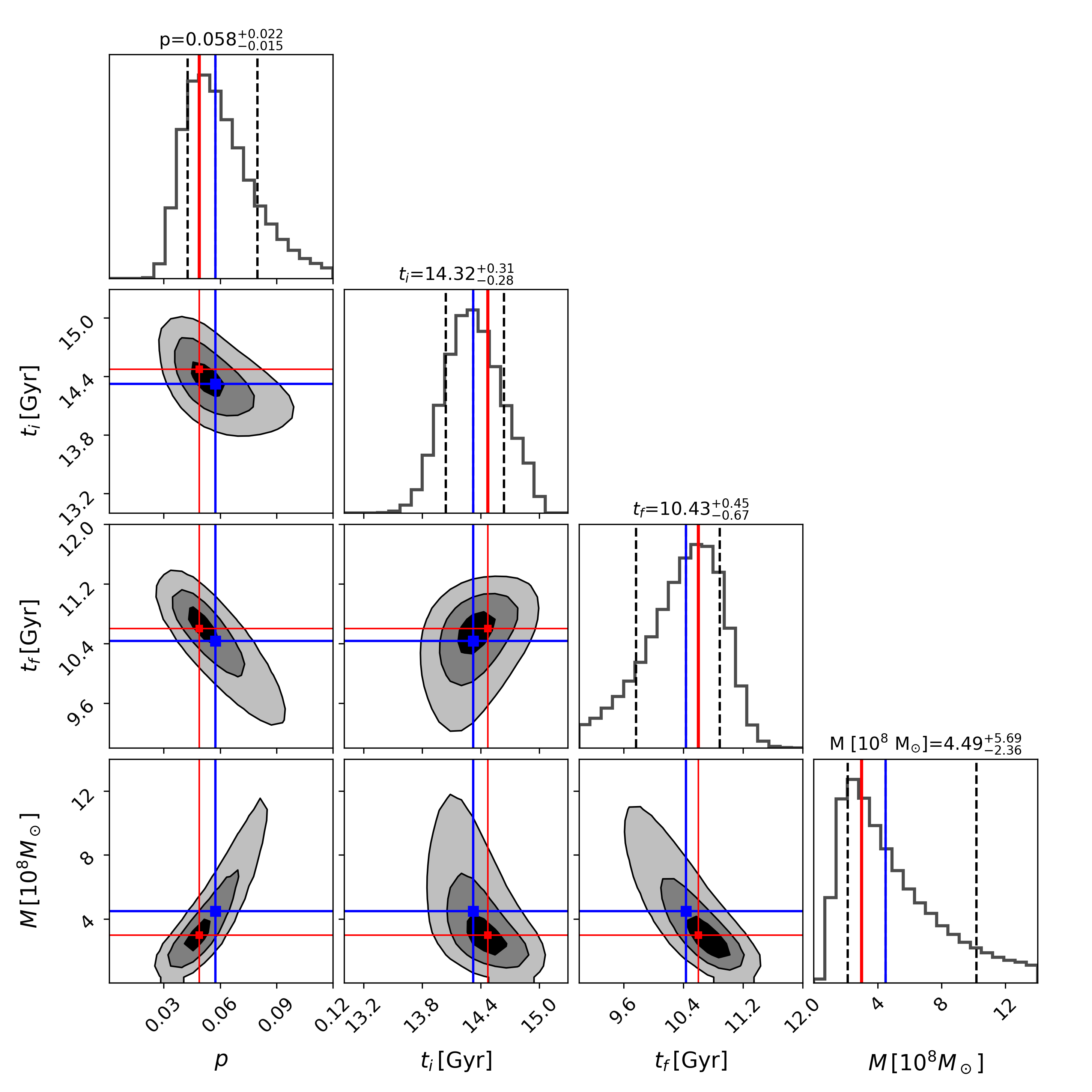}
    \caption{{\bf Corner plot of the GSE AMR model MCMC fit.} Corner plot of the joint posterior for the four global parameters, yield p, star formation initial time $t_i$, accretion time $t_f$, and derived mass M (in units of 10$^8$ M$_\odot$), for the GSE GCs. 
    Along the diagonal are the 1D histograms with dashed lines marking the 16th and 84th percentiles. Solid blue lines indicate the median and solid red lines the MAP, with titles giving the median ± uncertainties. Off-diagonal panels show the 2D posterior contours at the 16th, 50th and 84th percentile levels, with overplotted blue and red lines denoting the medians and MAP estimates for each parameter pair.
}
    \label{GSEcornerplot}
\end{figure}

The results of the AMR model fit are summarized in Supplementary Table~2, and interpreted in the main paper. The corresponding corner plots are shown in Fig.~\ref{cornerplot},~\ref{GSEcornerplot},~\ref{MWcornerplot}.\\

\vspace*{0.2cm}

\noindent{\bf 4. Historical accounting of accreted and in-situ populations in the inner Galaxy}\\

With our findings, we unequivocally demonstrate that a population of accreted tracers exists in the inner galaxy. Such a population bears chrono-dynamical properties that set it apart from the {\it in-situ} population even in the metal-poor regime, and is at the same time distinct from  other accreted components.

In this context, we are also in the position to clarify the confusion existing in the literature about the plethora of accreted substructures in the inner Galaxy, which can be summarized as follows.
Kruijssen et al. 2019\cite{kruijssen19} used a sample of GCs selected as accreted based on their location in the AMR \cite{marinfranch09, leaman13} to postulate the presence of three different merger events in the MW halo. These were deemed to be Sagittarius, Canis Major and a yet unknown one, named Kraken. Now that we have a better understanding of the MW assembly history, we know that while Sagittarius was obviously a correct identification, Canis Major does not exist \cite{momany06} and the majority of the GCs associated to it are actually members of GSE. The remaining clusters were used as an argument to predict the existence of the third merger, Kraken. We now know\cite{massari19, callingham22} that the GCs tentatively associated with Kraken also include members of GSE, Sequoia \cite{myeong19}, the Helmi Streams \cite{helmi99}, and the MW. Therefore, the original definition of Kraken is an (implausible) ensemble gathering everything that is neither Sagittarius nor GSE. Furthermore, using the EAGLE simulations, a mass of M$_{\star}\sim2\times10^9$ M$_{\odot}$ was estimated for the merger, significantly larger than what we have inferred here for LKH. The generous definition, coupled with the details of the comparison with the EAGLE cosmological simulations, might explain the discrepancy in the estimated masses.

Soon after Massari et al. 2019\cite{massari19} identified a group of about 20 GCs with coherent dynamical properties and which were not associated with any known merger (they also associated GCs with GSE, Sagittarius, Sequoia and the Helmi streams). They proposed that this ``Low-energy" group of GCs might have been the debris of an unknown accretion event. Kruijssen et al. 2020\cite{kruijssen20} adjusted the definition of Kraken to match that of this ``Low-energy" group. Since we demonstrate here that most of the ``Low-energy" group GCs are effectively tracing LKH, it is not surprising that the revised estimated properties\cite{kruijssen20} for Kraken are similar to the ones we find here for LKH. Although the accreted origin of the ``Low-energy" GCs, was supported later through other analyses\cite{callingham22}, it has been challenged by other works\cite{belokurov24, chen24} using  chemo-dynamical arguments. These works have argued that the enhanced [Al/Fe] of the GCs favours an in-situ origin, since  one might expect for stars from accreted galaxies\cite{horta21} that [Al/Fe] $< 0$. Noteworthy is that even if one removes the effect of  multiple populations which induce a scatter in [Al/Fe], the first  generation of stars in globular clusters associated with GSE 
also have [Al/Fe] $> 0$, as do the (extra-galactic) metal-poor globular clusters in the LMC\cite{mucciarelli09}. This demonstrates that [Al/Fe] should not be used\cite{vasini24} as an indicator of an accreted or {\it in-situ} origin for GCs.

Another structure that has been proposed to occupy the inner Galaxy is Heracles \cite{horta21}. The Heracles system is composed of stars with dynamical properties similar to the ``Low-energy" group GCs, and with chemical properties that confine them in the unevolved region of the [Mg/Mn] vs [Al/Fe] plane\cite{das20}. This chemo-dynamical selection of Heracles stars, although effective for [Fe/H]$>-1.6$, is not able to distinguish the accreted from the in-situ component in the metal-poor regime, which is the one probed by GCs in this work. Based on our findings, the progenitor characterized here enters the definition of Heracles, as part of the accreted population of tracers now located in the inner halo (within $\sim6$ kpc).
More precisely, the one described here is the cleanest definition of this merger event. As shown in many works\cite{horta24, wempe24, callingham25}, the early life of a galaxy like the Milky Way is characterized by many merger events, the majority of which of too low mass to host GCs. In this sense, by working with GCs as tracers, we are effectively filtering the progenitors to which we are sensitive in mass. As a result, our analysis is selecting a progenitor that is massive enough to host a system of GCs. The chemo-dynamically selected Heracles, instead, includes the entire population of early mergers experienced by the Milky Way. 

This interpretation agrees with the fact that the mass of LKH is nominally slightly lower than the mass of the Heracles ``composite population", though still consistent within the uncertainties\cite{horta21}. In fact, a lower limit on its total stellar mass, estimated by only considering unevolved stars with -1.5$<$[Fe/H]$<-1.0$, is M$_{\star}\simeq7.2\times10^8$ M$_{\odot}$\cite{horta25}. This is larger by about a factor of 1.5 compared to our estimate for LKH, but could be off by a larger factor given that the vast majority of LKH GCs have [M/H]$<-1.4$ ([Fe/H]$\lesssim-1.7$) and are not included in the mentioned Heracles selection.

Finally we note that if some of the LKH GCs (as well as those associated with the other progenitors) have been disrupted, this implies that the currently observed AMR is less-populated than it originally was, and that its original shape might have been slightly altered by this effect. However, this is unlikely to affect the clear separation of LKH's AMR from that of the other two massive progenitors that populate the inner Galaxy\cite{kruijssen19b}, although it might have a small impact on the associations of individual GCs to the various sequences.

In summary, the results of this work enable identifying LKH as a clean version, extended to low metallicity, of Heracles. Since LKH's constituents vastly match those associated with the Low-Energy group and Kraken, we can finally recognize LKH as a single massive structure accreted long ago, and that is now buried in the inner Galaxy.\\

\bmhead{Data availability}
Most of the GC photometric catalogs analysed here are available through the HUGS project portal  {\tt https://archive.stsci.edu/prepds/hugs/}. Those that are not present in that repository, will be shared upon request to D. Massari or R. Cohen.
GCs age estimates and dynamical associations are available at {\tt https://www.oas.inaf.it/en/research/m2-en/carma-en/}. 

\bmhead{Code availability}
The codes employed to perform the statistical analysis on the AMRs can be requested to C. Fanelli. The code used to perform the isochrone fit has been developed in the context of the CARMA\cite{massari23} and the {\it Hubble}-Missing Globular Cluster Survey\cite{massari25} projects, and will be released at the projects conclusion. Interested readers should contact D. Massari.

\bmhead{Acknowledgments}
The authors thank the referees for the thorough feedback that helped to significantly improve the paper. 
This paper should be considered as one of the contributions to the CARMA series of papers.
This work presents results from the European Space Agency (ESA) space mission Gaia. Gaia data are being processed by the Gaia Data Processing and Analysis Consortium (DPAC). Funding for the DPAC is provided by national institutions, in particular the institutions participating in the Gaia MultiLateral Agreement (MLA). 
Based on observations with the NASA/ESA HST, obtained
at the Space Telescope Science Institute, which is operated by
AURA, Inc., under NASA contract NAS 5-26555. 

\bmhead{Funding Statement} DM, SC, and EP acknowledge financial support from PRIN-MIUR-22: ``CHRONOS: adjusting the clock(s) to unveil the CHRONO-chemo-dynamical Structure of the Galaxy” (PI: S. Cassisi). DM acknowledges financial support by the Italian Ministry of University and Research (grant FIS2023$-$01611, CUP C53C25000300001). CF acknowledges support by the PRIN INAF 2023 grant ObFu N$^2$RED  (PI: C. Fanelli). CG and SC acknowledge support from the Agencia Estatal de Investigación del Ministerio de Ciencia e Innovación (AEI-MCINN) under grant ``At the forefront of Galactic Archaeology: evolution of the luminous and dark matter components of the Milky Way and Local Group dwarf galaxies in the Gaia era" with reference PID2020-118778GB-I00/10.13039/501100011033 and PID2023-150319NB-C21/C22/10.13039/501100011033. CG also acknowledge support from the Severo Ochoa program through CEX2019-000920-S.
M.M. acknowledges support from the Agencia Estatal de Investigaci\'on del Ministerio de Ciencia e Innovaci\'on (MCIN/AEI) under the grant ``RR Lyrae stars, a lighthouse to distant galaxies and early galaxy evolution" and the European Regional Development Fun (ERDF) with reference PID2021-127042OB-I00. AH, TC, EW, and HCW, acknowledge financial support from a Spinoza Award from NWO (SPI 78-411).
Co-funded by the European Union (ERC-2022-AdG, "StarDance: the non-canonical evolution of stars in clusters", Grant Agreement 101093572, PI: E. Pancino).

\bmhead{Author contributions}
DM has coordinated the project and led the scientific interpretation. CZ performed the photometric analysis and the isochrone fitting. CF has led the statistical analysis of the results. AH critically contributed to the interpretation of the results and to the writing of the manuscript. EC contributed to the photometric analysis. FAA developed the isochrone fitting code. AB produced the photometry of NGC4372. RC provided the photometric catalogs of NGC~6401, NGC~6254 and NGC~6256. TC and EW provided the analysis of cosmological simulations and contributed to the statistical analysis. HCW performed the GC orbital integration in a barred gravitational potential. SC, MM, CG, EP, SS, MS and AM are core members of the CARMA project.\\

\bmhead{Competing Interests Statement} There are no competing interests.\\

\vspace{1 cm}
\noindent\textbf{Additional information}

\noindent\textbf{Supplementary information} is available for this paper.

\noindent\textbf{Correspondence and requests for materials} should be addressed to D. Massari. 

\noindent\textbf{Author information} 
Davide Massari, https:/orcid.org/0000-0001-8892-4301

\newpage
\noindent\textbf{\large Supplementary Information} \\

\begin{table}[h!]
\caption{Results of isochrone fitting: for each cluster, values of [M/H], $E(B-V),$ distance modulus and age are reported along with their associated upper and lower limits.}
\label{tab:iso}
\renewcommand\arraystretch{1.5}
\[
\begin{array}{ccccc} 
\toprule
\hline
\text{Cluster Name} & \text{[M/H]} & \text{E(B-V)}  & \text{DM} & \text{Age}\\
                   &              &   \text{[mag]}  &  \text{[mag]} & \text{[Gyr]}\\
\midrule
\text{NGC4372} & -1.71^{+0.03}_{-0.04} & 0.48^{+0.01}_{-0.01} & 13.74^{+0.01}_{-0.01} & 14.48^{+0.12}_{-0.12}\\
\text{NGC4833} & -1.62^{+0.04}_{-0.06} & 0.33^{+0.01}_{-0.01} & 14.09^{+0.02}_{-0.02} & 13.43^{+0.26}_{-0.23}\\
\text{NGC5986} & -1.37^{+0.05}_{-0.05} & 0.29^{+0.01}_{-0.01} & 15.08^{+0.01}_{-0.01} & 13.61^{+0.14}_{-0.13}\\
\text{NGC6093} & -1.59^{+0.03}_{-0.02} & 0.21^{+0.01}_{-0.01} & 15.05^{+0.01}_{-0.01} & 13.72^{+0.06}_{-0.05}\\
\text{NGC6121} & -1.01^{+0.07}_{-0.05} & 0.43^{+0.01}_{-0.01} & 11.44^{+0.01}_{-0.02} & 12.24^{+0.22}_{-0.30}\\
\text{NGC6144} & -1.57^{+0.05}_{-0.03} & 0.44^{+0.01}_{-0.01} & 14.49^{+0.01}_{-0.01} & 14.09^{+0.14}_{-0.31}\\
\text{NGC6254} & -1.34^{+0.06}_{-0.08} &  0.26^{+0.01}_{-0.01} & 13.57^{+0.01}_{-0.01} & 13.71^{+0.42}_{-0.19}\\
\text{NGC6256} & -1.36^{+0.12}_{-0.08} &  1.11^{+0.01}_{-0.01} & 14.46^{+0.02}_{-0.01} & 13.31^{+0.32}_{-0.29}\\
\text{NGC6397} & -1.73^{+0.13}_{-0.10} & 0.17^{+0.01}_{-0.01} & 11.99^{+0.02}_{-0.02} & 14.17^{+0.51}_{-0.35}\\
\text{NGC6401} & -1.02^{+0.04}_{-0.03} & 0.93^{+0.01}_{-0.01} & 14.29^{+0.02}_{-0.02} & 13.66^{+0.39}_{-0.53}\\
\text{NGC6453} & -1.44^{+0.05}_{-0.06} & 0.64^{+0.01}_{-0.01} & 15.09^{+0.01}_{-0.01} & 13.54^{+0.26}_{-0.20}\\
\text{NGC6535} & -1.53^{+0.05}_{-0.07} & 0.42^{+0.01}_{-0.01} & 13.99^{+0.02}_{-0.02} & 13.55^{+0.32}_{-0.27}\\
\text{NGC6541} & -1.44^{+0.03}_{-0.04} & 0.11^{+0.01}_{-0.01} & 14.40^{+0.01}_{-0.01} & 13.31^{+0.22}_{-0.35}\\
\text{NGC6656} & -1.58^{+0.03}_{-0.03} & 0.36^{+0.01}_{-0.01} & 12.61^{+0.01}_{-0.01} & 14.40^{+0.25}_{-0.24}\\
\text{NGC6681} & -1.39^{+0.05}_{-0.07} & 0.10^{+0.01}_{-0.01} & 14.84^{+0.01}_{-0.01} & 13.78^{+0.27}_{-0.18}\\
\text{NGC6752} & -1.40^{+0.01}_{-0.01} & 0.06^{+0.01}_{-0.01} & 13.05^{+0.01}_{-0.01} & 14.30^{+0.15}_{-0.39}\\
\text{NGC6809} & -1.73^{+0.10}_{-0.03} & 0.19^{+0.01}_{-0.01} & 13.63^{+0.01}_{-0.01} & 13.30^{+0.18}_{-0.17}\\
\bottomrule
\end{array}
\]
\end{table}

\renewcommand{\arraystretch}{1.75} 
\begin{table}[!h]
\centering
\caption{Parameter estimates for LKH and GSE, with asymmetric $1\sigma$ errors.}
\label{tab:corner_systems}
\begin{tabular}{lcccc}
\toprule
System     & $p$                       & $t_i$           & $t_f$           & M$_{\star}$          \\[-0.5em]
     &                        & $[\mathrm{Gyr}]$           & $[\mathrm{Gyr}]$           & $[10^8\,M_\odot]$          \\
\midrule
 LKH      & $0.066^{+0.028}_{-0.019}$  & $14.52^{+0.27}_{-0.26}$       & $12.19^{+0.49}_{-0.59}$    & $6.30^{+8.95}_{-3.60}$       \\
 GSE                   & $0.058^{+0.022}_{-0.015}$  & $14.32^{+0.31}_{-0.28}$       & $10.43^{+0.65}_{-0.67}$    & $4.49^{+5.69}_{-2.36}$       \\
\bottomrule
\end{tabular}
\end{table}
\renewcommand{\arraystretch}{1.0} 

\end{document}